\documentclass[aps,twocolumn,apl,superscriptaddress,showpacs,reprint]{revtex4-1}

\usepackage{graphicx}
\usepackage{dcolumn}
\usepackage{bm }
\usepackage{upgreek}
\usepackage{paralist}
\usepackage{braket}
\usepackage{amsmath}
\usepackage{xcolor}
\usepackage{upgreek}

\newcommand{\Ca}{\ensuremath{^{40}{\rm Ca}^+\,}}

\begin{document}

\title{Quantum Sensing of Intermittent Stochastic Signals}
\author{Sara L. Mouradian}
\author{Neil Glikin}
\author{Eli Megidish}
\author{Kai-Isaak Ellers}
\author{Hartmut Haeffner}
\affiliation{Physics Department, University of California, Berkeley, California 94720, USA}
\affiliation{Challenge Institute for Quantum Computation, University of California, Berkeley, CA  94720}

\begin{abstract}

Realistic quantum sensors face a trade-off between the number of sensors measured in parallel and the control and readout fidelity ($F$) across the ensemble. We investigate how the number of sensors and fidelity affect sensitivity to continuous and intermittent signals. For continuous signals, we find that increasing the number of sensors by $1/F^2$ for $F<1$ always recovers the sensitivity achieved when $F=1$. However, when the signal is intermittent, more sensors are needed to recover the sensitivity achievable with one perfect quantum sensor. We also demonstrate the importance of near-unity control fidelity and readout at the quantum projection noise limit by estimating the frequency components of a stochastic, intermittent signal with a single trapped ion sensor. Quantum sensing has historically focused on large ensembles of sensors operated far from the standard quantum limit. The results presented in this manuscript show that this is insufficient for quantum sensing of intermittent signals and re-emphasizes the importance of the unique scaling of quantum projection noise near an eigenstate.
\end{abstract}


\maketitle

Quantum metrology aims to estimate a physical parameter of a signal via the response of a controllable quantum sensor coupled to the signal. Ideally, entanglement between quantum sensors can be exploited to break classical sensing limits~\cite{Giovannetti2011Apr}, but even without entanglement, quantum systems can reach the quantum projection noise (QPN) limit~\cite{Itano1993May} --- a noise floor unattainable by classical systems --- and provide high sensitivity and precision~\cite{Degen2017}. The figure of merit for a quantum sensor depends on the application. In this manuscript we focus on the minimum parameter that can be accurately resolved, or the sensitivity. The sensitivity depends on the response of the sensor's state to the parameter of interest and on the noise on the measurement of the sensor's state. 

Ideally, a quantum sensor would be made of a large ensemble of $M$ individual quantum systems with a long ensemble coherence time and unity fidelity state preparation, control, and measurement across the ensemble. However, any experimental implementation must balance the gain in sensitivity due to increased $M$ with any potential loss in fidelity due to increased decoherence and non-uniform control~\cite{Barry2020Mar}. Amplitude and frequency estimation of coherent signals has been well studied~\cite{Boss2017May,Glenn2018,Kristen2020Jun}, and some work had been done on spectroscopy of stochastic signals~\cite{Bylander2011May,Laraoui2010,Young2012,Meriles2010,Taylor2008Oct,Norris2016Apr,Yuge2011Oct,Alvarez2011Nov}. In all of these studies, the signals are continuous. However, some signals of interest are intermittent~\cite{Lyne2009, Gray2020Aug} with durations much shorter than standard coherence times. Sensing protocols for intermittent signals must differ significantly from sensing of continuous signals because the signal duration --- not the sensor's coherence time --- limits the integration time for each individual measurement. A recent paper discussed quantum sensing of a specific intermittent signal~\cite{Gefen2019}. In this manuscript, we extend the theory to highlight experimentally relevant features of quantum sensing of intermittent signals, and provide an experimental demonstration.

 \begin{figure}[hbtp]
    \centering
    \includegraphics[scale=1]{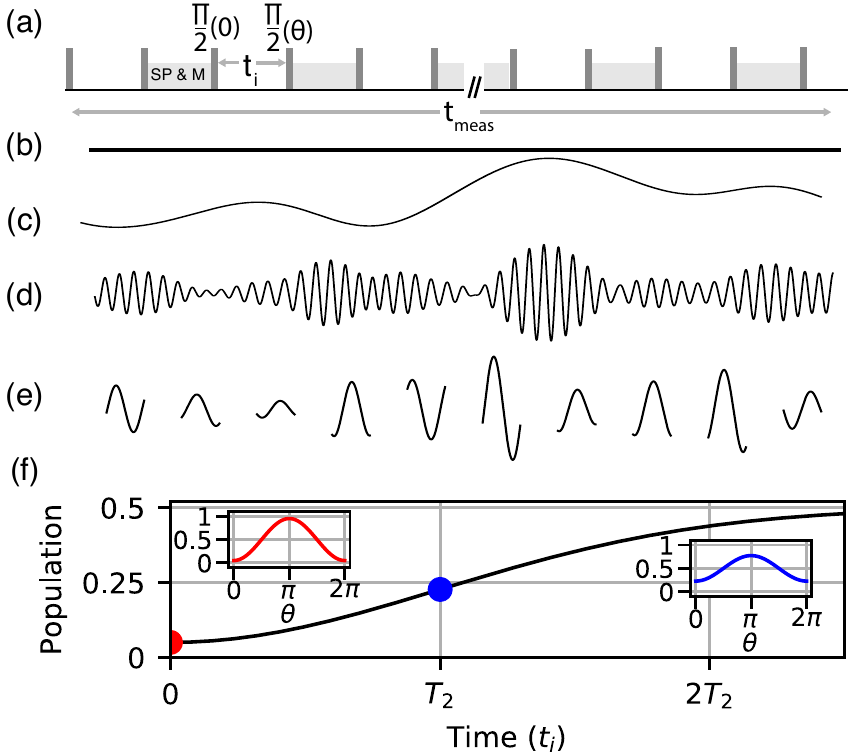}
    \caption{(a) Depiction of the Ramsey sequence used for sensing including the time needed for state preparation and measurement (SP and M). The phase ($\theta$) of the second $\pi/2$ pulse is chosen to optimize the sensitivity. (b-e) The signals considered in this manuscript: (b) A constant amplitude signal. (c) A stochastic signal. (d) A two-frequency stochastic signal. (e) An intermittent two-frequency stochastic signal. (f) A Ramsey signal with increasing $t_i$ with no signal applied with population $p = 0.5(1-C(t))$. Insets show the reduced contrast at $t=0$ and at the $T_2$ time of the sensor.}
    \label{signals}
\end{figure}


We investigate how a sensor's sensitivity scales with number of sensors $M$ and their fidelity $F$ for four scenarios of interest illustrated in Fig.~\ref{signals}(b-e): (1) Amplitude estimation of a constant signal (Sec.~\ref{sec:AmpCoh}); (2) Variance estimation of the amplitude of a 0-mean stochastic signal (Sec.~\ref{sec:AmpStoh}); (3) Frequency estimation of a two-frequency stochastic signal (Sec.~\ref{sec:FreqStoh}); and (4) Frequency estimation of an {\it intermittent} two-frequency stochastic signal (Sec.~\ref{sec:FreqInt}). The stochastic signals are all normally distributed and have an autocorrelation time longer than each individual Ramsey integration time, ($t_i$), but shorter than the total measurement time ($t_{\text meas}$) comprising $N$ individual integration periods plus the additional time needed for state preparation and measurement (see Fig.~\ref{signals}(a)).


These signals cause a small change $B(g,t)$ in the transition frequency of the sensor. We estimate this change, and thus the parameter of interest, $g$, via a Ramsey measurement as depicted in Fig.~\ref{signals}(a)~\cite{Degen2017}. $N$ projective measurements of the population ($p$) of $M$ sensors are used to estimate the average phase ($\phi(g,t_i)$) accrued by the sensor due to $B(g,t)$ during an integration time $t_i$. The interferometer is biased with a controllable phase $\theta$ to optimize the response to the parameter of interest. The $M$ sensors are unentangled such that the noise on the population measurement is bounded below by the QPN, $\sigma^2_{\text{qpn}} = p(1-p)/(NM)$.

Errors in state-preparation, control, and measurement contribute to a time-independent sub-unity fidelity $F$. Decoherence due to coupling to the environment reduces the maximum achievable contrast over time: $C(t) = Fe^{-\chi(t)}$ for a known function $\chi(t)$. Here, we assume that the coherence of the sensor is limited by slow noise and the contrast exhibits a Gaussian decay characterized by a coherence time $T_2$, $\chi(t) = \frac{t^2}{2T_2^2}$. Fig.~\ref{signals}(f) plots $p(0,t) = 0.5(1-C(t)\cos(\theta))$ for a sensor with $F=0.9$ and an arbitrary $T_2$ time. Insets show the dependence on the bias phase $\theta$ at $t_i = 0$ and $t_i = T_2$. For $t_i=0$, the contrast is dominated by sub-unity state preparation, operations, and measurement. For $t_i = T_2$ the loss in contrast is dominated by decoherence.

With the field applied, the sensor accrues a field-dependent phase, and the population is $p(g,t) = \frac{1}{2}\left[1-C(t)\cos(\theta + \phi(g,t)) \right]$. The sensor's signal is then the change in final population, $\Delta p(g,t) = p(g,t) - p(0,t)$. We assume $p(0,t)$ is known exactly, and the noise is bounded by the QPN at $p(g,t)$ but may include readout and preparation noise if the sensor does not achieve the QPN limit.

The sensitivity is the parameter $g_{\text{min}}$ for which the signal-to-noise ratio (SNR) is equal to 1~\cite{Degen2017}. For each of the signals above we consider the effect of $F$ on sensitivity and find the number of sensors with $F < 1$ that are needed to recover the sensitivity achievable with a single $F=1$ sensor.


\section{Amplitude Estimation}
\label{sec:Amp}
\subsection{A Coherent Signal}
\label{sec:AmpCoh}

\begin{figure}[hbtp]
    \centering
    \includegraphics[scale=1]{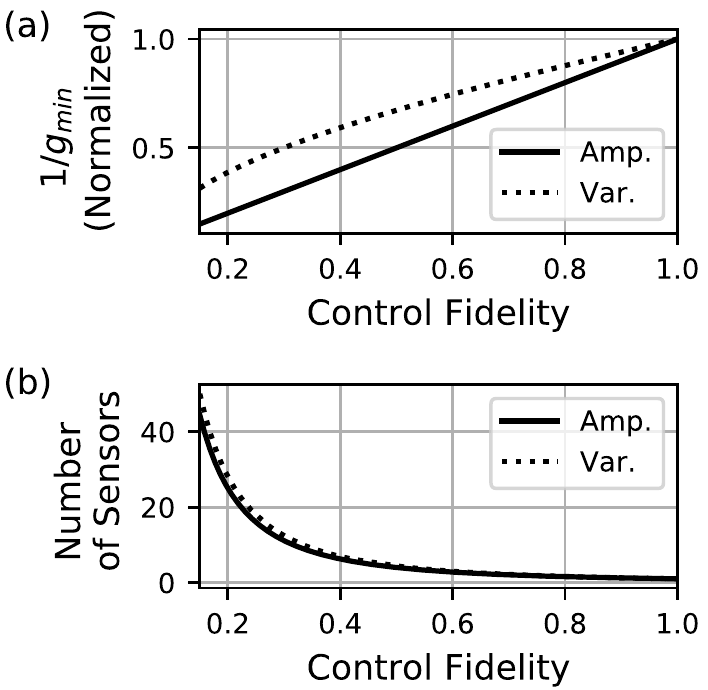}
    \caption{Comparison between amplitude (solid) and variance (dotted) estimation. (a) The effect of sensor fidelity on the achievable sensitivity for a fixed number of sensors. (b) The number of sensors needed to achieve the same sensitivity as a single unity-fidelity sensor.}
    \label{ampComparison}
\end{figure}

We begin with the well-studied case of amplitude estimation of a signal $B(g, t) = g$, such that $\phi(g,t) = \int_{0}^{t_i} B(g,t)dt = gt_i$. The optimal SNR is obtained by biasing the measurement at $\theta = \pi/2$ so that $p(0,t) = 0.5$~\cite{Degen2017}. Small $g$ will give a linear sensor response $\Delta p(g,t) \approx \frac{1}{2}C(t) g t_i$ and the QPN is unaffected by the presence of the signal. In this case, the minimum detectable $g$ is 
\begin{equation}
    g_{\text{min}} = (\sqrt{NM} t_i C(t_i) )^{-1}.
\end{equation}
which is minimized at a Ramsey integration time of $t_i = T_2$. Fig.~\ref{ampComparison} shows the effect of $F<1$. Non-unity fidelity reduces the achievable sensitivity, but $M = 1/F^2$ sensors recovers the sensitivity of one $F=1$ sensor. This is a familiar result~\cite{Degen2017}, and state-of-the-art amplitude estimation demonstrations leverage large ensembles to improve the sensitivity even while sacrificing contrast~\cite{Barry2020Mar} as large ensembles of quantum sensors are found naturally in solid state systems~\cite{Clevenson2015Apr} or atomic vapor cells~\cite{Budker2007Apr}. 

\subsection{A Stochastic Signal}
\label{sec:AmpStoh}
Now we consider variance estimation of a slowly varying,  normally distributed, stochastic signal $B_s$ with a mean $\left< B_s \right> = 0$ and variance $\left< B_s^2 \right> = g^2$. The average population of the sensor with the signal applied is
\begin{eqnarray}
\label{eq:popVar}
\left<p(g,t_i)\right> &=&  \frac{1}{2}\left( 1 - C(t_i)\left<\cos(\theta + B_s t_i)\right> \right) \\ 
&=& \frac{1}{2}\left( 1 - C(t_i))\cos(\theta) e^{-g^2t_i^2/2} \right).
\end{eqnarray}
The signal adds an additional effective source of decoherence, $1/T_{2,\text{eff}}^2 = 1/T_2^2 + g^2 $. Thus, it is advantageous to bias the measurement at $\theta = 0,\pi$ such that $p(0,t_i)$ measures the full contrast without the signal applied. In this way, $p(g,t_i)$ reflects any reduction in contrast due to $g$. Then,
\begin{eqnarray}
\Delta p(g,t_i) &=& \frac{1}{2}C(t_i)(1-e^{-g^2t_i^2/2})\\
& \simeq & \frac{1}{2}C(t_i)g^2t_i^2/2\\
\sigma_{\text{qpn}}^2 & = & \frac{1}{4NM}(1-C(t_i)^2 + 2C(t_i)^2g^2t_i^2 )\\
\text{SNR} = \frac{\Delta p}{\sigma_{\text qpn}} &= &\frac{\sqrt{NM} C(t_i) g^2t_i^2}{2\sqrt{1-C^2(t_i) + C^2(t_i)g^2t_i^2}}
\end{eqnarray}
for $(g t_i)^2 \ll 1$. Setting SNR=1 and solving for $g^2$, we find
\begin{equation}
\label{eq:gminVar}
    g_{\text{min}}^2 = \frac{2C(t_i) + 2\sqrt{C(t_i)^2 + (1-NMC(t_i)^2)}}{NMC(t_i)t_i^2}.
\end{equation}
which is minimized at $t_i \sim \sqrt{2}T_2$. As shown in Fig.~\ref{ampComparison}(a), the scaling of sensitivity with fidelity for variance estimation ($ \sim 1/\sqrt{F}$) differs from the scaling for amplitude estimation ($1/F$).  However, the scalings with $M$ for variance and amplitude estimation ($\sim 1/\sqrt[4]{M}$, $1/\sqrt{M}$ respectively) change in step such that $M\simeq1/F^2$ sensors still nearly compensate for $F<1$ as seen in Fig.~\ref{ampComparison}(b). Thus, when building a quantum sensor for variance detection it is again best to increase $M$ if the subsequent decrease in $F$ is more modest than $1/\sqrt{M}$.

\section{Frequency Estimation}
\label{sec:Freq}
\begin{figure}[hbtp]
    \centering
    \includegraphics[scale=0.95]{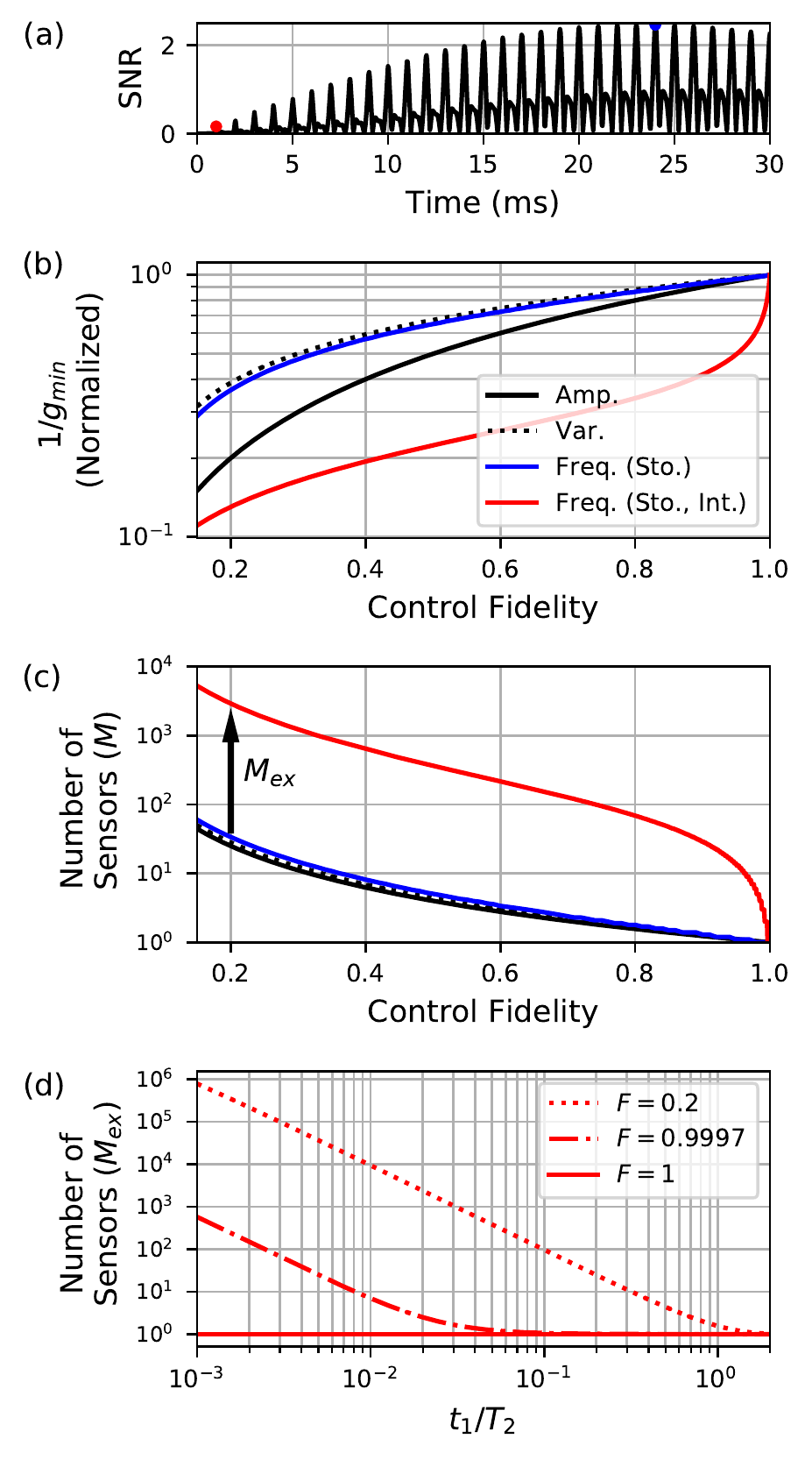}
    \caption{Frequency estimation of an intermittent stochastic signal as described by Eq.~\ref{eq:freqSig} with $\omega_s=2\pi\times1$~kHz, $\sigma= 2\pi\times 500$~Hz, $N=1000$, $M=1$. (a) Time dependence of the SNR for frequency estimation of a $g=2\pi\times10$~Hz signal. (b) The effect of sensor fidelity on the achievable sensitivity for amplitude and variance estimation (black solid, dotted) and frequency estimation of stochastic continuous (blue) and intermittent (red) signals. (c) The number of sensors needed to achieve the same sensitivity as single unity-fidelity sensor. The color scheme is the same as in (b). (d) The number of sensors $M_{\text ex}$ needed to achieve the same sensitivity as that for a continuous signal. As the signal time decreases it is increasingly difficult to compensate for $F<1$ by increasing $M$.}
    \label{fig:freqSto}
\end{figure}

In this section, we consider frequency resolution of a two-frequency stochastic signal (Fig.~\ref{signals}(d)) with a finite auto-correlation time that is longer than $t_i$ but much shorter than $t_{\text{meas}}$~\cite{Gefen2019}. The signal has the form
\begin{equation}
\label{eq:freqSig}
    B_{\text{fe}}(t) = A_1\sin\omega_1 t + B_1\cos\omega_1 t + A_2\sin\omega_2 t + B_2\cos\omega_2 t.
\end{equation}
We describe this signal with three parameters: the frequency separation $g = \omega_1 - \omega_2$, the center frequency $\omega_s = (\omega_1 + \omega_2)/2$, and $\sigma^2$, the variance of the 0-mean normal distribution that describes all four amplitudes $A_{1,2}$, $B_{1,2}$. We assume prior knowledge of $\omega_s$ and $\sigma$ and estimate the frequency separation, $g$. If $\omega_s$ and $\sigma$ are not previously known, multivariate estimation techniques can be used~\cite{Gefen2019,Demkowicz-Dobrzanski2020Apr}. 

A similar estimation problem has been considered in the spatial domain~\cite{Lupo2016,Tsang2016}, and sub-Rayleigh discrimination of the position of two incoherent light sources has been demonstrated~\cite{Tham2017Feb,Nair2016}. This method has also been used in conjunction with sum-frequency generation to achieve discrimination between optical frequency and temporal modes~\cite{Donohue2018Aug}.

\subsection{A Stochastic Signal}
\label{sec:FreqStoh}
For a continuous signal, we consider the problem numerically. We simulate the response to the signal described in Eq.~\ref{eq:freqSig} with $\omega_s=2\pi\times1$~kHz and $\sigma= 2\pi\times 500$~Hz with $N=1000$ measurements of 1 sensor. These parameters are relevant for our experimental implementation. As with variance estimation, we measure an additional time-dependent decoherence due to the signal so the measurement should be biased at $\theta = 0,\pi$ to measure the full contrast.

In Fig.~\ref{fig:freqSto}(a) we plot SNR$(t_i)$ for $g=2\pi\times10$~Hz which is locally maximized at integer multiples of the period of the center frequency, $t_n = 2\pi n / \omega_s$ with an optimal SNR near $\sqrt{2}T_2$ (blue marker in Fig.~\ref{fig:freqSto}(a)). In Fig.~\ref{fig:freqSto}(b,c) we see that frequency estimation of a stochastic signal scales similarly to variance estimation with respect to $M$ and $F$. Again, it is optimal to prioritize increasing $M$ over increasing $F$ if the subsequent decrease in $F$ scales more favorably than $1/\sqrt{M}$.

\subsection{An Intermittent, Stochastic Signal}
\label{sec:FreqInt}
Finally, we consider an intermittent, stochastic two-frequency signal as in Fig.~\ref{signals}(e). In contrast to Sec.~\ref{sec:FreqStoh}, the intermittent signal has individual durations $t_{\text{sig}} < T_2$, such that each integration time is limited by the signal duration and not the sensor's coherence.  The signal for frequency estimation approaches 0 for short measurement times, but a recent proposal~\cite{Gefen2019} claims that the sensitivity of a perfect quantum sensor is unaffected by integration time because the QPN also approaches 0 near an eigenstate. As in Sec.~\ref{sec:AmpStoh},\ref{sec:FreqStoh}, we bias the measurement to measure the decrease in contrast due to the signal. However, unlike the previous sections, $t_i < T_2$ so the sensor is near an eigenstate as illustrated in Fig.~\ref{signals}(f), and it is possible to take full advantage of the fact that QPN approaches 0 for $p=0,1$ and recover a finite sensitivity to $g$~\cite{Gefen2019}.

We consider a signal which only exists for bursts of durations less than $2\pi \times 2/\omega_s$ so that it is optimal to measure at $t_1 = 2\pi/\omega_s$ (red marker in Fig.~\ref{fig:freqSto}(a)). For longer signals, the optimal sensing procedure is discussed in Ref.~\cite{Gefen2019}. At $t_1$ we solve for the sensitivity analytically. The phase accrued by the sensor due to the signal is $\phi(g) = \int_0^{t_1} B_{\text{fe}}(t)dt$. For $g \ll \omega_s$, the average population of the sensor with the signal applied is
\begin{eqnarray}
\left<p(g)\right> &=&  \frac{1}{2}\left( 1 - C_t\left<\cos(\phi(g)\right> \right) \\ 
&\simeq&  \frac{1}{2}\left(1-C_t e^{-\frac{4\pi^2\sigma^2}{\omega_s^4}g^2}\right)
\label{eq:popInt}
\end{eqnarray}
where $C_t=C(t_1)$. The derivation follows the steps in Sec.~\ref{sec:AmpStoh}, and finding the SNR and solving for $g$ with SNR=1, we find
\begin{equation}
\label{eq:wrminInt}
    g_{\text{min}} = \frac{\omega_s^2}{2\pi\sigma}
    \sqrt{\frac{C_t + \sqrt{C_t^2 + NM(1-C_t^2)}}{NMC_t}}.
\end{equation}

The red curve in Fig.~\ref{fig:freqSto}(b) shows a strong departure from previous scalings of sensitivity with $F$. This arises because the QPN decreases rapidly towards 0 as the system approaches an eigenstate, and the sensor can only be near an eigenstate if $F\sim1$. Thus, for $F\sim1$, the scaling of the sensor's sensitivity deviates strongly from the $\sim\sqrt{F}$ scaling found for the sensitivity for frequency or variance estimation of continuous stochastic signals. As a result, significantly more sensors are necessary to recover the sensitivity achieved with a single unity fidelity sensor as seen in Fig.~\ref{fig:freqSto}(c). 

For low $F$, $M\sim(1/F^2) (1/(1-e^{-2\chi(t_1)}))$ which has the same scaling as the previous cases, but with a factor $M_{\text ex}$ which increases rapidly for $t_1 \ll T_2 $. This extra scaling with signal duration is shown in Fig.~\ref{fig:freqSto}(d) for a perfect sensor ($F=1$), a sensor with the best demonstrated readout and control fidelity to date ($F=0.9997$)~\cite{Christensen2020Apr}, and a low-fidelity sensor ($F=0.2$). For a sensor with unity fidelity, there is no dependence on the signal duration as reported by Ref.~\cite{Gefen2019}. However, for any $F<1$, as the duration of the stochastic signal ($t_1$) decreases, the number of sensors $M_{\text ex}$ needed to achieve the same sensitivity as that for a continuous signal increases: $M_{\text ex} \propto 1/t_1^2$ for $F\ll C(t_1)$. For instance, if $t_1 / T_2 = 10^{-3}$, $M \sim 10^6 (1/F^2)$ for $F\ll1$. Thus, when building a sensor for intermittent signals it is necessary to maintain $F$ and remain at the QPN limit while increasing $M$.

\section{Experimental Demonstration}
\label{sec:Experiment}
Finally, we present the first experimental demonstration of quantum sensing of an intermittent signal. Our quantum sensor is a \Ca ion. Specifically, we use the electronic ground state $\ket{S} \equiv\,^{2}$S$_{1/2}$, and the long-lived metastable state, $\ket{D} \equiv\,^2$D$_{5/2}$ in a Ramsey interferometry measurement. A single measurement of the sensor's state is performed via resonant excitation of a cycling state-dependent transition. The number of photons detected during the detection time is thresholded to infer whether the detection event corresponded to the ion in the $\ket{S}$ (bright) or $\ket{D}$ (dark) state. This is repeated over $N$ measurements to obtain a measurement of $\left<p\right>$. 

\begin{figure}[hbtp]
    \centering
    \includegraphics[scale=0.9]{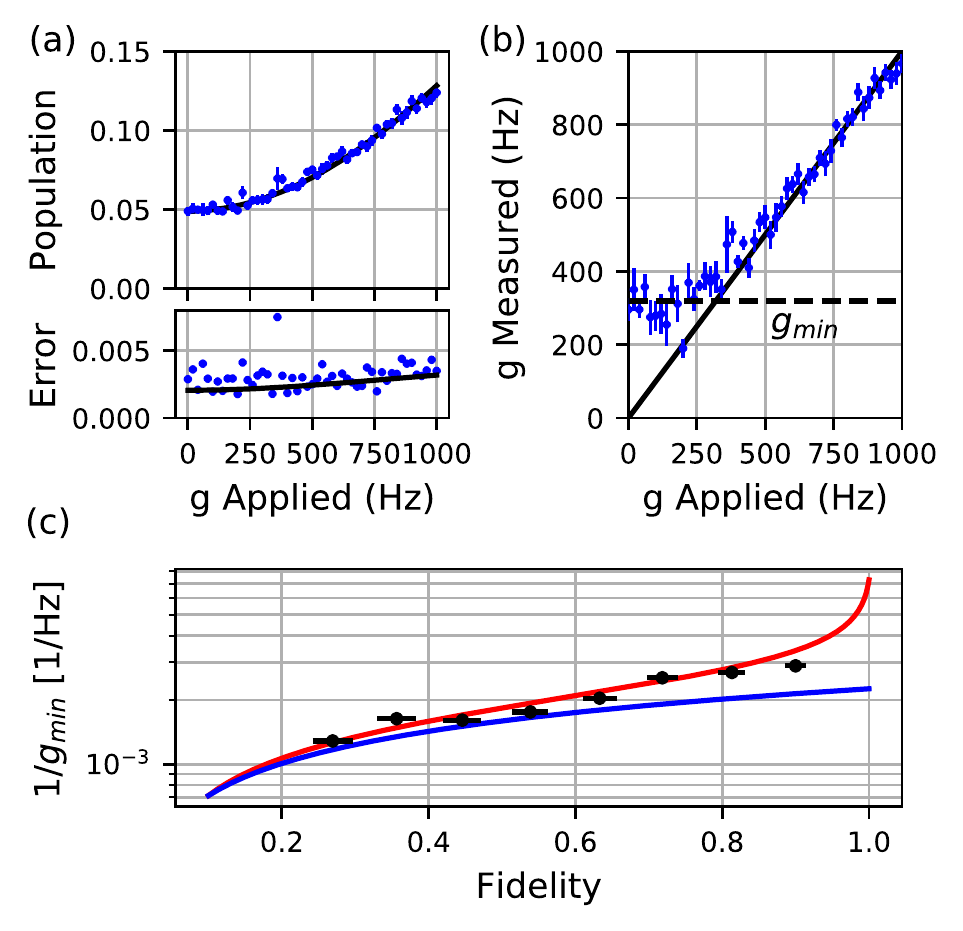}
    \caption{(a) Population as a function of 
    frequency separation for $\omega_s = 2\pi \times 2$~kHz, $\sigma=2\pi \times 275$~Hz and the error over 11 sets of $N=1000$ measurements ($M=1$) compared to the error expected from QPN. (b) Estimation of the applied frequency separation, $g$ from the population data. (c) Comparison between the experimentally derived sensitivity and the theoretical scalings for intermittent (red) and continuous (blue) stochastic signals.}
    \label{fig:expSetup}
\end{figure}

We implement the intermittent signal in Eq.~\ref{eq:freqSig} with an ac stark shift on the $\ket{0}$ to $\ket{1}$ sensing transition. We program an arbitrary waveform generator with $N$ instances of Eq.~\ref{eq:freqSig} with $N$ different values of the amplitudes $A_{1,2},B_{1,2}$ pulled from a 0-mean normal distribution with variance $\sigma^2$ as described in Sec.~\ref{sec:Freq}. The output of this arbitrary waveform generator modulates an rf source driving an acousto-optical modulator controlling the amplitude of laser light detuned 250~kHz from the sensing transition and focused onto the ion. The amplitude of the signal is limited by the available optical power and achievable modulation depth. This ac stark shift signal necessarily has a dc bias. We use a Hahn-Echo sequence to remove the effect of this dc bias and any slowly varying environmental noise. However, the signal is on only during the first half such that the sensing protocol remains an un-echoed Ramsey sequence.  A phase scan at $g=0$ bounds the contrast ($C = 0.903 \pm 0.015$) at the given measurement parameters due to the sensor decoherence ($T_2 = 7.97 \pm 0.51$~ms) and control fidelity ($F = 0.91 \pm 0.015$). The phase scan also gives the measurement phase $\theta$ to bias the sensor as close as possible to an eigenstate as described in Sec.~\ref{sec:FreqInt}.

We consider the response of our trapped ion sensor to varying frequency separations $g$ for a signal with $\omega_s = 2\pi \times 2$~kHz and  $\sigma = 2\pi \times 275$~Hz. Fig.~\ref{fig:expSetup}(a) shows the increase in population (or decrease in contrast) as $g$ increases. The error is the standard error of 11 repetitions of $N=1000$ measurements of $M=1$ sensors. A comparison between the standard error over those 11 repetitions and the error expected from QPN (Fig.~\ref{fig:expSetup}(a), bottom panel) shows that our noise is nearly dominated by QPN, with 17\% excess uncorrelated noise at $g=0$.

We estimate $g$ using Eq.~\ref{eq:popInt} for each of the 11 measurements of $p(g)$. For $g < g_{\text{min}}$, nearly half of the measurements fall below $p(g=0)= 0.047$. The estimation of $g$ via Eq.~\ref{eq:popInt} is undefined for these measurements, and we disregard them. Fig.~\ref{fig:expSetup}(b) shows the results of the frequency estimation. For $g > g_{\text{min}}$, the estimated $g$ matches the applied value, while the estimation is biased for $g < g_{\text{min}}$ due to measurements below $p(g=0)$. We use this to extract an experimentally derived $g_{\text{min}}$. We verify this method with numerical simulations which fit the analytical expression of Eq.~\ref{eq:wrminInt}. For an intermittent, stochastic 2~kHz signal with $t_i =0.5$~ms and a total measurement time of 500~ms we expect to achieve a 290~Hz sensitivity. We measure 320~Hz sensitivity. This discrepancy is due to the noise above the QPN limit and underlines the importance of reaching the QPN limit in the sensing of intermittent signals. 

Finally, in order to demonstrate the strong dependence on sensor fidelity laid out in Sec.~\ref{sec:FreqInt}, we post-process our data to artificially reduce the effective contrast of our detection and re-derive an empirical $g_{\text{min}}$. As discussed above, for each of the $N$ measurements done at every data point, we record either a 0 or 1 depending on the photon counts we detect. To reduce contrast, we record the wrong value some percentage of the time. We do this with the data taken at $g=0$ as well to get an accurate measurement of the fidelity. In Fig.~\ref{fig:expSetup}(c) we find that the scaling with $F$ matches our expectations (red), and clearly deviates from the $1/\sqrt{F}$ scaling expected for a continuous signal (blue).  

\section{Conclusion}
We have compared variance and frequency estimation of stochastic signals to the familiar case of amplitude detection. We find that for continuous signals, sub-unity control fidelity is easily compensated for with a modest ensemble of sensors. However, if the signal is intermittent, near-unity control fidelity and measurements at the QPN limit are significantly more important and the number of sensors needed to recover the same sensitivity increases significantly. This manuscript has focused on a specific intermittent signal, but similar results are expected for more general intermittent signals. Experimental implementations of quantum sensing have historically focused on increasing the number of sensors in an ensemble. However, here we show that this is not sufficient if the integration time is limited by the signal and not the sensor's coherence time. Thus it is important to continue optimizing the control fidelity of quantum sensors and to achieve the standard quantum limit when adding more sensors. 

\begin{acknowledgments}
We thank Alex Retzker and Soonwon Choi for valuable conversations during the preparation of this manuscript. S.M. was supported by an appointment to the Intelligence Community Postdoctoral Research Fellowship Program at University of California, Berkeley, administered by Oak Ridge Institute for Science and Education through an interagency agreement between the U.S. Department of Energy and the Office of the Director of National Intelligence. This work has been supported by the ARO MURI grant W911NF-18-1-0218. The apparatus has been supported by ONR through Grant No. N00014-17-1-2278 and by the NSF Grant No. PHY 1620838. 
\end{acknowledgments}

\bibliography{SensingPaper_v7.bib}

\end{document}